\newcommand{\ale}{\ \raisebox{-.3ex}{$\stackrel{<}{\scriptstyle \sim}$}\ }
\newcommand{\age}{\ \raisebox{-.3ex}{$\stackrel{>}{\scriptstyle \sim}$}\ }

\documentstyle{mn}

\input psfig

\title[Discs in ejected T Tauri stars]{The ejection of T Tauri stars 
	from molecular clouds and the fate of circumstellar discs}
	
\author[P.J. Armitage and C.J. Clarke]{P.J. Armitage\thanks{Present 
	address: Canadian Institute 
	for Theoretical Astrophysics, McLennan Labs, 60 St George Street,
	Toronto M5S 1A7, Ontario, Canada} and C.J. Clarke \\
 	$$ Institute of Astronomy, Madingley Road, Cambridge, CB3 0HA}	
 	
\begin{document}

\maketitle

\begin{abstract} 	
We investigate the evolution of circumstellar discs around T Tauri stars 
that are ejected from small stellar clusters within molecular clouds. 
In particular, we study
how the interaction that leads to ejection may hasten the transition
between Classical and Weak-lined T Tauri status. In our models, 
ejections of T Tauri stars at velocities of 3-10 km/s truncate the
accretion disc at radii between 1 and 10 a.u., reducing the viscous
evolution time of the disc so that accretion rapidly ceases. The
observational appearance of the resulting systems is then dependent
on the presence or absence of a stellar magnetic field. For
non-magnetic stars we find that a near-infra red excess should 
persist due to reprocessing of stellar radiation, but that this
is greatly diminished for magnetic T Tauri stars by the presence
of a magnetosphere extending to corotation. In either case, there
is a period when ejected stars should appear as non-accreting 
systems with detectable circumstellar material at wavelengths
of 5 microns and beyond. We discuss the implications of these
results for models in which ejected stars contribute to the 
halo of pre-main-sequence objects discovered from ROSAT observations
of star forming regions and the All-Sky Survey.

\end{abstract}


\section{Introduction}

	The ROSAT All-Sky Survey (RASS) has identified large
numbers of candidate Weak-lined T Tauri stars (WTTS) in and
around nearby star-forming regions, including Taurus-Auriga
(Neuh\"auser et al. 1995a), Chamaeleon (Alcal\'a et al.
1995) and Orion (Alcal\'a 1994; Sterzik et al. 1995). 
Spectroscopic follow-up of subsamples of these candidates 
finds strong Li absorption in around 60\% of surveyed sources 
(Alcal\'a et al. 1995, 1996), and in some cases provides
kinematic evidence that the stars newly identified from
the RASS have radial velocities consistent with the
previously-known WTTS population (Neuh\"auser et al. 1995b).  
In most case, however, such evidence is lacking, and
indeed the most striking observation is the large
spatial extent of the candidate WTTS, many of which lie
far ($10^\circ$ or more) from known star-forming regions
(Montmerle \& Casanova 1995; Neuh\"auser et al. 1995b;
Sterzik et al. 1995). 

	The distances of the newly identified sources
are mostly highly uncertain, making age determinations
from the position of stars in the Hertzprung-Russell diagram
suspect. Montmerle \& Casanova (1995) quote very young ages
($\simeq 10^5$ yr) for some sources relatively close
(a few degrees) to Chamaeleon, while Neuh\"auser et al.
(1995b) give upper limits of 25 Myr for a sample of WTTS
south of Taurus based on Li abundances. For a larger
sample of ROSAT discovered sources, Magazzu et al. (1996)
find an overabundance of Li as compared
to the Pleiades for a range of spectral types, including
those (eg. K and later) for which the Li width becomes
important as an age indicator. These
results suggest that the selection criteria for the
larger samples {\em are} finding a new population of WTTS
younger than the Pleiades, though contamination from
much older ($10^8$ yr) stars may also be important
(Hartmann, private communication), especially for
the purpose of statistical studies.

Here, we note that if any significant fraction
of the newly identified WTTS are young (i.e. have ages
in the 1-10 Myr range common for previously known WTTS),
then their origin so far from known sites of star formation
poses two severe problems. Firstly, how did these sources
arrive at their current location, and secondly, why is
a corresponding population of dispersed {\em Classical} T Tauri
stars not also observed.
One possibility is that the number
of stars forming in small isolated groups has previously
been underestimated, as suggested by Feigelson (1996) on
the basis of simple kinematic models for the distribution
of the dispersed RASS stars. Alternatively, the halo
sources might have been formed within known regions of
star formation activity, before being ejected by dynamical
interactions. In this latter scenario the number and 
observed distribution of the outliers implies that a 
reasonable fraction of all low-mass stars must be ejected
from molecular clouds at relatively high velocities of
several ${\rm kms}^{-1}$ or greater.

	The decay of small stellar clusters has been considered
as a contributor to the halo of ROSAT sources by Sterzik \&
Durisen (1995), and in the context of binary formation by
McDonald \& Clarke (1995). 
In these calculations decay of the cluster and ejection of most
of its members occurs rapidly -- within a few tens of crossing 
times (a few $\times 10^5$ - $10^6$ yr for the McDonald \& Clarke
simulations, and earlier for the denser clusters modelled by
Sterzik \& Durisen). Although the lifetime of circumstellar discs
varies widely (Strom 1995), at such early epochs most of the
ejected stars would be expected to be Classical T Tauri
stars (CTTS), surrounded by circumstellar discs. Since CTTS are 
generally well-localised in dark clouds (eg. Alcal\'a et al. 1995),
a viable ejection model for the halo ROSAT sources requires that
the ejected CTTS are converted into exclusively Weak-lined
systems (where there is little or no evidence for circumstellar
material) on a timescale that is short compared to the ages of
the halo sources. 

        In this paper, we discuss the effects of an encounter leading to
ejection on the accretion disc of a CTTS. The ejection of stars with
velocities of a few ${\rm kms}^{-1}$ implies interactions 
that are close enough to
severely curtail the extent of the accretion disc, but not so close as to
destroy it entirely. The weakened disc will then evolve more rapidly
because of the reduced viscous timescale at its outer edge, so that the
rate of accretion onto the star falls rapidly. However significant
K excesses could still arise from reprocessing of the stellar luminosity by
a passive disc, and we show that the strength of this contribution to 
the spectral energy distribution depends critically on the magnetic
field of the star. For weakly magnetic stars (where the disc extends
to the stellar equator) the passive flux from the disc will persist
long after active accretion has ceased, whereas a strong stellar
field can disrupt the hot inner disc regions and reduce the reprocessing
flux coincident with the decline in the mass accretion rate.
Magnetic fields of the required strength have been directly detected in
several WTTS (Basri, Marcy \& Valenta 1992; Guenther \& Emerson
1995, 1996), and many indirect lines of evidence
suggest that they are common in both WTTS and CTTS (see, eg, the review by
Hartmann 1994).

        The plan of this paper is as follows. In Section 2, we estimate
the accretion rates below which magnetic and non-magnetic systems 
would appear as WTTS on the basis of their near-infrared colours
and $H \alpha$ equivalent width. In Section 3 we use these
estimates to calculate the timescale for the CTTS $\rightarrow$ WTTS
transition, and compute the stellar densities required to eject
a large number of T Tauri stars from their parent clouds.
In Section 4 we verify these timescale estimates
by directly computing the time-dependent evolution of the star-disc system 
for a typical choice of parameters. Section 5 summarizes our findings and
discusses the observational consequences of the model.

\section{Models for T Tauri accretion discs}

\begin{figure}
 \psfig{figure=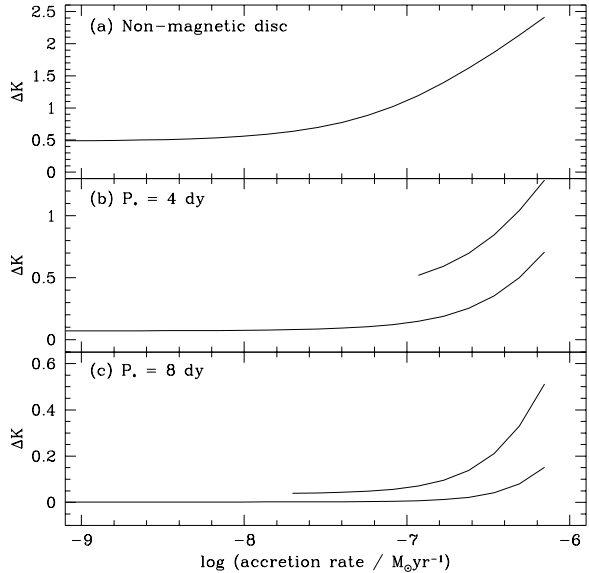,width=3.25truein,height=3.25truein}
 \caption{K excesses over the stellar photosphere for a variety of
 				accretion disc models described in the text, as a function 
 				of the steady-state
 				accretion rate through the disc. (a) A non-magnetic disc
 				extending to the stellar surface. (b) A magnetically 
 				disrupted disc around a star of period $P_* = 4 \ \rm{dy}$
 				with $R_{\rm m} = R_{\rm c}$ (lower curve) and
 				$R_{\rm m} = 0.7 \times R_{\rm c}$ (upper curve).
 				(c) A magnetically disrupted disc around a star of period 
 				$P_* = 8 \ \rm{dy}$ with $R_{\rm m} = R_{\rm c}$ (lower curve) 
 				and $R_{\rm m} = 0.7 \times R_{\rm c}$ (upper curve).
 				The curves for magnetic discs with $R_{\rm m} = 0.7 \times
 				R_{\rm c}$ do not extend to very low $\dot{M}$, since at
 				low accretion rates the magnetosphere must lie at larger
 				radii.}
 \label{fig1}
\end{figure} 

The parameters and mechanisms that control the transition of a star from 
Classical to Weak-Line status are largely unknown, but it is natural to
assume that the mass accretion rate through the disc $\dot{M}$ plays
an important role. With this assumption, we define three mass fluxes;
the accretion rate for a typical CTTS, $\dot{M}_{\rm CTTS}$, the accretion
rate below which a star appears as a WTTS in a non-magnetic model
$\dot{M}_{\rm weak, no B}$, and the corresponding quantity in a
magnetospheric model $\dot{M}_{\rm weak, B}$. A variety of
observational characteristics distinguish WTTS from CTTS (see, eg.
the review by Bertout 1989), but low equivalent width of $H \alpha$
($\ale 10 {\rm \AA}$) and lack of an infra-red excess at K ($2.2 \ \mu{\rm m}$)
are common identifying features. We adopt these conditions
for deciding when model systems have become WTTS, and 
estimate the values of $\dot{M}_{\rm weak, no B}$ and 
$\dot{M}_{\rm weak, B}$. These quantities are then used to 
estimate the timescale for the CTTS to WTTS conversion in magnetic
and non-magnetic systems that suffer an ejection interaction.

Figure 1 shows the excess flux $\Delta K$ at $2.2 \ \mu{\rm m}$
for a range of discs around a star of mass $1 \ M_\odot$
and radius $2 \ R_\odot$. The models include the heating
of the disc from accretion, reprocessing of stellar radiation, and
where appropriate work done by magnetic torques 
(assuming a stellar dipole field of $1 \ {\rm kG}$). 
The models are calculated
using the formalism outlined in the Appendix. No contribution
from a boundary layer or magnetically funnelled accretion shock is
included as these components of the total system luminosity are
not expected to be strong at K (note that in magnetospheric 
models the infalling material is unlikely to be optically
thick at K even for $\dot{M} = \dot{M}_{\rm CTTS}$, and will
certainly have $\tau_{\rm K} \ll 1$ for the lower accretion
rates of interest here). For the magnetic models we have
assumed that the disc is disrupted inside some magnetospheric
radius $R_{\rm m}$, which is taken either to be at corotation 
$R_{\rm c}$ (where the Keplerian disc has angular velocity 
$\Omega = \Omega_*$), or at $0.7 \times R_{\rm c}$. 
$R_{\rm m} \approx R_{\rm c}$ is expected on theoretical
grounds at low $\dot{M}$ (eg Armitage \& Clarke 1996; 
see also Wang 1996), while
fitting of magnetic disc models to the spectral energy distributions
of CTTS typically gives $R_{\rm m} \approx 0.7 \times R_{\rm c}$
(Kenyon, Yi \& Hartmann 1996). Additional
angular momentum loss via stellar winds (e.g. Tout \& Pringle 1992)
would be expected to reduce $R_{\rm m}$ below the values seen
in models that assume magnetic disc linkage as the only
braking mechanism. The stellar rotation rate $P_*$
is taken to be 4 dy or 8 dy, these values being
typical for WTTS and CTTS respectively (Bouvier et al. 1995).

For the non-magnetic disc model, reprocessing of the stellar luminosity 
ensures that an easily detectable K excess is found for all accretion 
rates, provided only that the disc remains optically thick at K. 
This will be the case down to very low accretion rates. To
order of magnitude, an optical depth $\tau_{\rm K} = 1$ requires
a disc column density $\Sigma = 0.1 \ {\rm g/cm^2}$, which is
at least 4 orders of magnitude below the surface density seen in
CTTS disc models accreting at $10^{-7} \ M_\odot \ {\rm yr^{-1}}$. 
The accretion rate corresponding to $\Sigma = 0.1$ depends on the
disc viscosity $\nu$ via $\dot{M} \sim 3 \pi \nu \Sigma$, where
$\nu$ is typically an increasing function of $\Sigma$. We conclude
that $\dot{M}_{\rm weak, no B}$ is likely to be more
than 4 orders of magnitude below $\dot{M}_{\rm CTTS}$, and that
a disc around a non-magnetic T Tauri star would remain visible 
via its K excess long after significant accretion (evidenced by
high equivalent width of $H \alpha$) had ceased.  

For the magnetic models, the behaviour of $\Delta K$ as a function
of $\dot{M}$ is very different. If $R_{\rm m} = 0.7 \times R_{\rm c}$,
then significant K excesses of $\sim 1 \ {\rm mag}$ are obtained at
accretion rates of $10^{-7} \ M_\odot {\rm yr^{-1}}$ and above for
$P_* = 4 \ {\rm dy}$, with smaller but still readily detectable 
$\Delta K$ values for the longer rotation period. However, once the
accretion rate has fallen sufficiently low that the magnetosphere 
lies at corotation, the models predict that $\Delta K$ should be
low. The dependence of $R_{\rm m}$ on $\dot{M}$ is (Clarke et al. 1995),
\begin{equation}
  R_{\rm m} = \left({{2B_*^2 R_*^6}\over {\dot{M}\sqrt{GM_*}}}\right)^{2/7}.
\label{2.1}
\end{equation} 
Thus, taking as observationally determined the result that
$R_{\rm m} \approx 0.7 \times R_{\rm c}$ in the Classical
T Tauri phase, we find that the magnetosphere
should lie at corotation once $\dot{M} \ale 0.3 \times \dot{M}_{\rm CTTS}$. 
From the Figure, we then conclude that T Tauri systems 
rotating at $P_* = 8 \ {\rm dy}$ would appear as
WTTS on the basis of their K excess below an accretion rate
of a few $\times 10^{-8} \ M_\odot {\rm yr^{-1}}$. For the
shorter rotation
period of 4~dy, $\Delta K$ falls below 0.1 magnitudes at approximately
the same $\dot{M}$.

These results imply that judged by the near infra-red excess, a 
magnetic CTTS would appear indistinguishable from a WTTS at an accretion rate 
of $\sim 10^{-8} \ M_\odot \ {\rm yr^{-1}}$ or below. 
The equivalent width of $H \alpha$ in
WTTS shows considerable scatter (Alcal\'a et al. 1993), 
so accretion at approximately this
level probably cannot be excluded on the basis of the $H \alpha$
measurements either. Hence, we estimate that $\dot{M}_{\rm weak, B}
\sim 10^{-8} \ M_\odot \ {\rm yr^{-1}}$, and conclude that
$\dot{M}_{\rm weak, B} \gg \dot{M}_{\rm weak, no B}$. This
critical accretion rate is then used in the following Section to 
estimate the timescale on which interactions leading to 
ejection from a cluster will reduce  $\dot{M}$ below 
$\dot{M}_{\rm weak, B}$, and hence convert a CTTS to Weak-lined
status.
 
\section{Timescale for the CTTS $\rightarrow$ WTTS transition}

The decay of small stellar clusters has been studied by
a number of authors using N-body simulations (van Albada
1968; Sterzik \& Durisen 1995), and techniques that attempt
to incorporate additionally the effects of discs (McDonald \&
Clarke 1995) or gas (Bonnell et al. 1996). Such clusters
are found to dissolve within a few tens of crossing times,
leading to the ejection of most of the members and the
formation of one central binary. The median velocities of the
escapers $v_{\rm eject}$ follow the scalings established 
for the 3-body problem,
\begin{equation}
 \left< v_{\rm eject} \right> \approx 0.5 \times \left(
 {|E_0|} \over {\left<m_{\rm eject}\right>} \right)^{1/2},
\label{2.11}
\end{equation}
where $m_{\rm eject}$ are the masses of the ejected stars,
and $|E_0| \propto M_{\rm c}^2 / R_{\rm c}$ is the total
energy of the system with mass $M_{\rm c}$ and scale length
$R_{\rm c}$ (Valtonen \& Mikkola 1992; Sterzik \& Durisen
1995). For a small cluster containing $N$ stars of mass
$M_*$, the above expression implies,
\begin{equation}
 R_{\rm c} \sim 10^3 \ {\rm a.u.} \times \left( N \over 10 \right)
 \left( M_* \over M_\odot \right) \left( v_{\rm eject} \over
 {5 \ {\rm km/s}} \right).
 \label{2.12}
\end{equation}  
This broadly agrees with the numerical results suggesting
that stellar separations in the 300-1000 a.u. range are required
to generate significant numbers of escapers with $v_{\rm eject}$
greater than a few ${\rm kms}^{-1}$. The required clusters are
thus vastly denser than the stellar density in regions
such as Taurus-Auriga, even when account is given to the
observed degree of subclustering (Gomez et al. 1993). However
such clusters are also short lived, typically $< 10^5$ yr 
(Sterzik \& Durisen 1995), and hence the observed state
of Taurus {\em cannot} necessarily be taken as evidence that
stars were not born in much more compact groups. Indeed
the dissolution of such initial clusters might plausibly
lead within a few Myr to {\em both} high velocity escapers and 
dilute aggregates (composed of stars ejected at only 
$\sim$ a ${\rm kms}^{-1})$ similar to those observed by 
Gomez et al. (1993). Detailed population synthesis would
be required to address this possibility in a more
quantitative fashion. As discussed by McDonald \& Clarke
(1995) and Sterzik \& Durisen (1995), the formation of
stars within such transitory clusters is consistent both
with some theoretical models of star formation (Boss 1993),
and the observed high fraction of binaries in the pre-main-sequence
(eg. Ghez 1996). 

For a dynamical ejection model to be relevant to the
formation of the RASS halo of WTTS, $v_{\rm eject}$ must 
be a few ${\rm kms}^{-1}$ or greater. Here we assume 
that conditions in the initial stellar clusters are
conducive to producing such velocities (i.e. similar
to those used as initial conditions by Sterzik \& 
Durisen 1995), and calculate the effects of those 
ejections on circumstellar discs.

To relate $v_{\rm eject}$ to the distance of
closest approach in the encounter that led to the ejection, 
$R_{\rm peri}$, we make use of prior work on
the scattering of single stars by binaries 
(for details see, e.g. Davies 1995). The
energy lost from the binary $\Delta E$ is related to the
binding energy of the binary, $E_{\rm bin}$, via
\begin{equation}
 \Delta E \ale 0.4 \times E_{\rm bin} = 
 0.4 \times {{G M^2} \over {2 d_{\rm bin}} },
\label{2.15}
\end{equation} 
where $d_{\rm bin}$ is the separation of the binary before the
encounter. Making use of the further result that
$d_{\rm bin} \approx R_{\rm peri}$, we find 
\begin{equation}
 v_{\rm eject} 
    \approx 15 \ {\rm km/s} \times \left( R_{\rm peri} \over {1 \ {\rm a.u.}}
    \right)^{-1/2},
\label{2.2}
\end{equation}
for stars of solar mass. Ejection velocities of
$v_{\rm eject} = 3 - 10 \ {\rm km/s}$, then imply $R_{\rm peri} \sim
2 - 25 \ {\rm a.u.}$. Star-disk encounters have been extensively
modelled (Hall, Clarke \& Pringle 1996; Clarke \& Pringle 1993),
and these simulations suggest that the accretion disc following
the encounter has an outer edge at $\sim R_{\rm peri} / 2$. The
ejected stars would thus be expected to harbour remnant discs
with outer radii $R_{\rm out} \ale 10 \ {\rm a.u.}$ immediately
following the interaction.

The weakened disc will then evolve {\em faster} as a consequence
of the reduced viscous timescale $t_\nu$ at $R_{\rm out}$. The
viscous timescale is given by (e.g. Pringle 1981),
\begin{equation}
 t_\nu \sim { 1 \over {\alpha \Omega} } \left( R \over H \right)^2,
\label{2.3}
\end{equation}
where $\alpha$ is the Shakura-Sunyaev viscosity parameter, $H$ is
the disc scale height, and all quantities are evaluated at 
radius $R$. For stars of solar mass,
\begin{equation}
 t_\nu \sim 1.6 \times 10^4 \ {\rm yr} \left( \alpha \over 10^{-3} 
 \right)^{-1} \left( (H/ R)_{R_{\rm out}} \over 0.1 \right)^{-2}
 \left( R_{\rm out} \over {1 \ {\rm a.u.}} \right)^{3/2},
\label{2.4}
\end{equation}
where $\alpha$ is the Shakura-Sunyaev viscosity parameter
appropriate to the outer regions of the disc following the
interaction.

Thus provided that $\alpha_{R_{\rm out}} \age 10^{-3}$ (as is likely
since even $\alpha=10^{-3}$ implies worryingly large disc masses
at radii of several a.u.) and $(H/R) \age 0.1$, the viscous 
timescale of the truncated disc will be short compared to 
the typical disc lifetime in CTTS. For example at 5 a.u., which
corresponds to the outer edge of the disc in a star ejected
at 5 ${\rm kms}^{-1}$, $t_\nu \sim 2 \times 10^5$ yr with 
the fiducial parameters given above. The
same conclusion follows by noting that unless $\alpha$ 
or $(H/R)$ are strongly increasing functions of $R$ in the
outer regions of the disc, the radial
dependence of $t_\nu$ is approximately as $\Omega^{-1} \propto
R^{3/2}$. A typical T Tauri disc has a lifetime of a few Myr, which 
implies $t_\nu \sim 1 \ {\rm Myr}$. Reducing $R_{\rm out}$ by a 
modest factor of $\sim 5$ then ensures that immediately 
following an encounter (before the disc has had time to 
expand viscously) the viscous timescale will be reduced
by around an order of magnitude. Estimates of T Tauri
disc sizes are generally $\sim 10^2$ a.u. or greater
(McCaughrean \& O'Dell 1996), and hence encounters
at radii of $\sim 10$ a.u. are likely to meet this
condition.

The viscous timescale can be regarded as a characteristic time
for the decay of the disc surface density and accretion rate.
As $\dot{M}_{\rm CTTS} \approx 10^{-7} \ M_\odot \ {\rm yr}^{-1}$,
only a few viscous times will be required before $\dot{M}$ through
the truncated disc falls below $\dot{M}_{\rm weak, B}$.
Thus within a few times $t_\nu$ (equation \ref{2.4}), the initially 
Classical T Tauri star will lose the $\Delta K$ and $H \alpha$ 
signatures of a CTTS, and will appear as a Weak-Lined system. 
In this model circumstellar material will still exist around the star
for a further, perhaps lengthy, period, but the
low accretion rate and the hole created by the stellar magnetosphere
renders it undetectable in the near infra-red.

We emphasise that the same conclusion does {\em not} follow for
non-magnetic disc models. In this scenario, the interaction reduces
$t_\nu$ exactly as for the magnetic case, but a large number
of viscous times are still necessary before $\dot{M}$ falls
below $\dot{M}_{\rm weak, no B}$. Before the accretion rate
falls this low, the disc will have had time to re-expand to
its original extent, restoring $t_\nu$ to a much larger value. 
As a consequence, the reduction in the duration of the
CTTS phase will be much less marked, and we would expect
the ejected systems to display infra-red excesses from
passive discs without significant active accretion.

\section{Numerical simulations}

\begin{figure*}
 \hspace{-.25truein}
 \psfig{figure=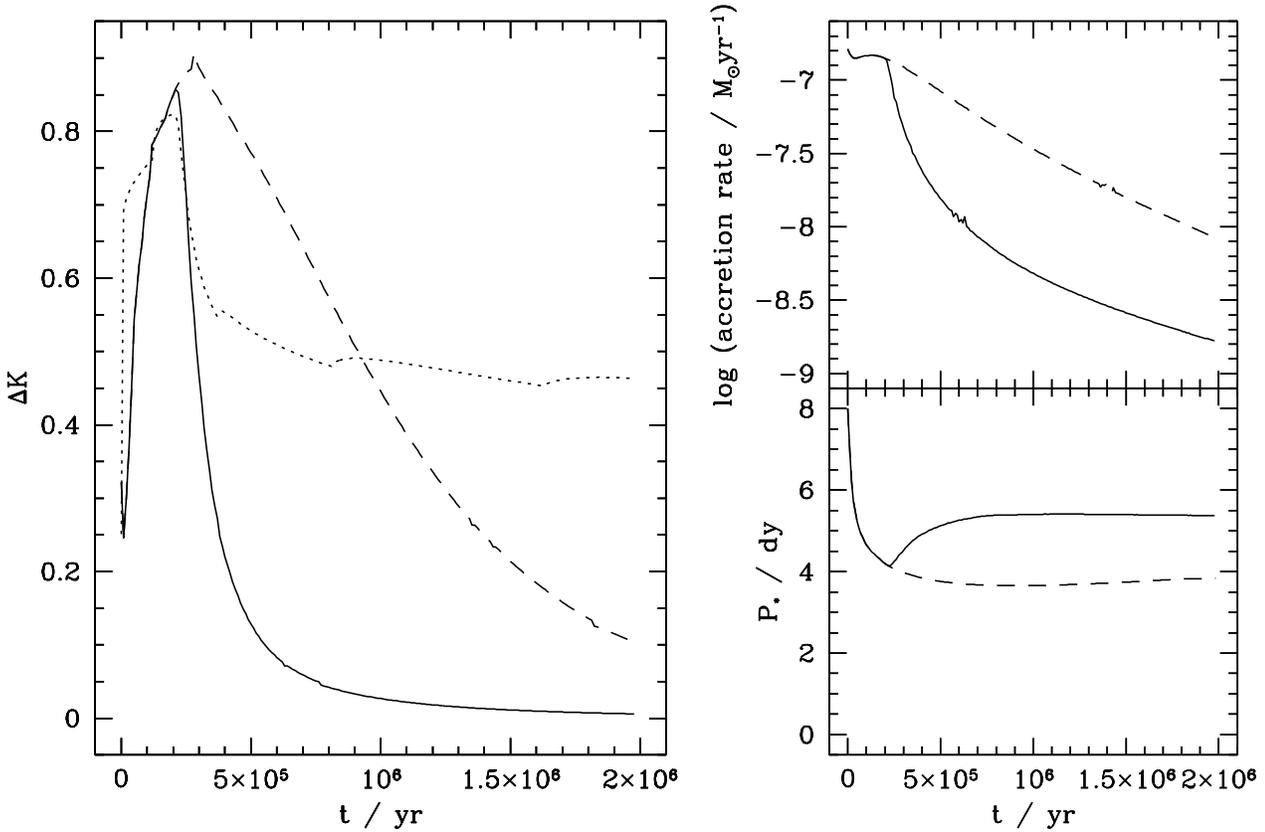,width=7truein,height=7truein}
 \vspace{-2.2truein}
 \caption{Comparison of the evolution of an undisturbed star-disc
 				system (dashed line) with one experiencing an encounter
 				at $t = 2 \times 10^{5} \ \rm{yr}$ that removes all
 				mass outward of 5 a.u. (solid curve). Both models
 				assume a disc disrupted at small radii by a stellar
 				magnetosphere.
 				The panels show the time-dependence of the
 			   monochromatic flux excess $\Delta K$ at $2.2 \ \mu{\rm m}$,
 			   accretion rate, and stellar spin period. 
 			   The K excess for a non-magnetic system undergoing the
 			   same encounter as the magnetic model is also shown
 			   (dotted line).}
 \label{fig2}
\end{figure*} 				

To verify the estimates derived in Section 2, we have calculated 
explicitly the evolution of the coupled star-disc system following
an encounter that destroys the outer accretion disc. The response
of the disc to magnetic and viscous torques is described using
the evolution equation given by Livio \& Pringle (1992),
\begin{eqnarray}
 {\partial \Sigma \over{\partial t}}= 
 {3 \over R} {\partial \over{\partial R}}
 \left[ R^{1/2} {\partial \over {\partial R}} (\nu \Sigma R^{1/2}) \right] 
 \nonumber \\
 + {1 \over R} {\partial \over {\partial R}} \left( {{\Omega - \Omega_*}
 \over \Omega} {{B_z^2 R^{5/2}} \over {\pi \sqrt{GM_*}}} \right),
\label{3.1}
\end{eqnarray}
which we solve together with equations for the spin period and magnetic
field of the star. The stellar model used is a 0.9985 $M_\odot$ model 
calculated using the most recent version of the Eggleton code 
(Pols et al. 1995), as modified for pre-main-sequence evolution by 
Tout (private communication). The model and numerical details are
described elsewhere (Armitage \& Clarke 1996). 

Figure 2 shows the time-dependence of the disc mass, accretion rate,
stellar spin period and K excess for two models; one in which the
disc beyond 5~a.u. is lost at $t = 2 \times 10^5 \ {\rm yr}$
(representing the effect of an ejection interaction), and a `control'
model in which the disc evolves undisturbed. 
We also show the evolution of $\Delta K$ for
a non-magnetic model subject to the same disruptive encounter.
The main parameters
of the simulations are; an outer boundary condition of $\Sigma = 0$
at 50 a.u., a stellar magnetic field $B_* = 1500 \ G \times
(P_* / 4 {\rm dy})$, an initial disc mass of $0.3 \ M_\odot$, and
an initial accretion rate of $1.5 \times 10^{-7} \ M_\odot {\rm yr}^{-1}$.
The viscosity prescription assumes an $\alpha$ in the inner disc
of $10^{-3}$, and gives an evolutionary timescale for the unperturbed
disc of $\sim 10^6 \ {\rm yr}$, consistent with the
inferred lifetime of CTTS discs (Simon \& Prato 1995).
The K excess $\Delta K$ is calculated using identical models to
those of the previous Section, except that the assumption that
the disc is in a steady state is relaxed and we use instead the
computed surface density to obtain the active accretion component.

Following the encounter, the accretion rate drops rapidly, reflecting
the reduced viscous timescale at the new outer edge ($t_\nu$ decreases
by a factor $\sim 10$ following the interaction). At late times, when the 
weakened disc has re-expanded to its previous size, the accretion
rate is a factor of $\sim 6$ times lower than in the model that
evolves undisturbed, or approximately
$(1-2)\times 10^{-9} \ M_\odot {\rm yr}^{-1}$
at $t = 2 \ {\rm Myr}$ for these models. The rotation period of
the `ejected' star is marginally increased as compared to the control 
model, because the spin-up torque from accretion is reduced while the
spin-down torque from magnetic braking is unaffected. The K excess
drops rapidly due to a combination of lower $\dot M$, increased
ratio of $R_{\rm m} / R_{\rm c}$ (which is $\sim 1$ in the
weakened disc simulation), and increased $R_{\rm c}$ due to the
slower rotation. $\Delta K$ falls below 0.1 magnitudes within
0.5 Myr, and is negligible at the end of the simulation.
This reduction in $\Delta K$ as compared to the control run
(which is itself still magnetically disrupted at small $R$)
amounts to a factor of $\sim 10$, in agreement with the estimates
presented previously. The run in which a non-magnetic disc is
subjected to an encounter also follows the expected behaviour. 
Immediately after the interaction, $\Delta K$ drops as for
the magnetic case due to the reduced $\dot{M}$. However, at
late times the K excess remains strong due to the reprocessing 
luminosity of the undisrupted disc at small radii.

\begin{figure*}
\psfig{figure=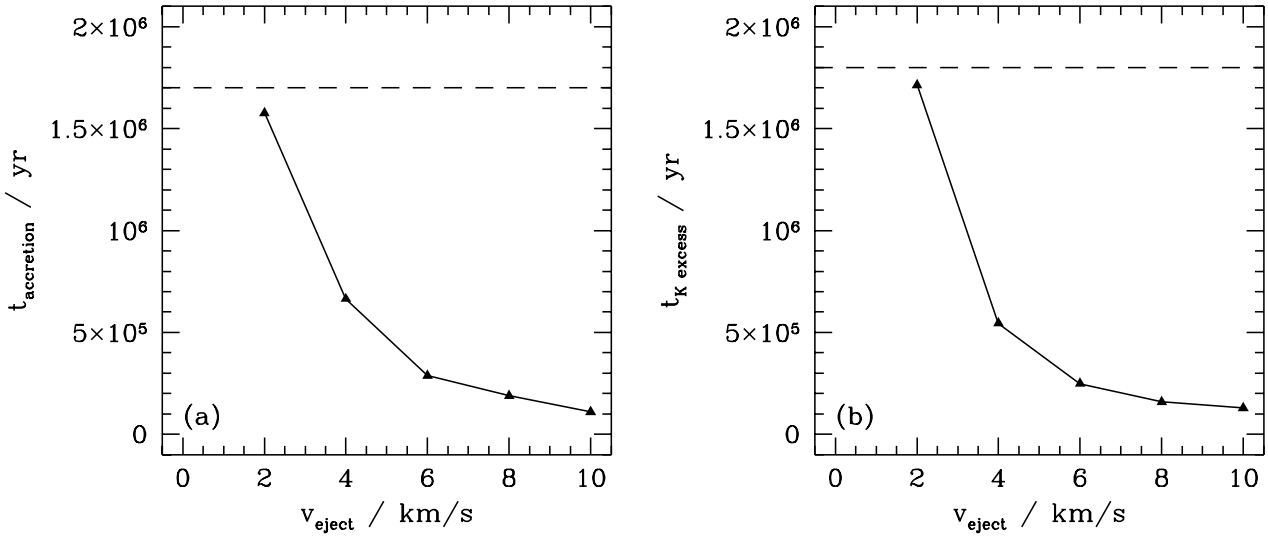,width=7.0truein,height=7.0truein}
\vspace{-3.5truein}
\caption{Effect of the encounter on the accretion rate and
 K excess, as a function of the ejection velocity $v_{\rm eject}$ 
 (or, equivalently, the periastron separation).
 (a) Time after encounter required for the accretion rate to
 fall below $10^{-8} \ M_\odot {\rm yr}^{-1}$. (b) Time
 required for the K excess to drop below 0.1 magnitudes.
 In both panels the period required for the undisrupted 
 disc model to reach these limits is shown as a dashed line.}
\label{fig3}
\end{figure*}

The extent to which the length of the CTTS phase is reduced by 
encounters is shown as Figure 3. A series of calculations with
the parameters given above were run, varying only the outer
radius at which the disc was truncated by the interaction.
The time subsequent to the interaction required for $\dot{M}$
to fall to $10^{-8} \ M_\odot {\rm yr}^{-1}$ is plotted as
a function of $v_{\rm eject}$, which is related to $R_{\rm peri}$
via equation (\ref{2.2}). Also plotted is the time required for
$\Delta K$ to fall below 0.1 magnitudes.

From the Figure, it can be seen that encounters that produce
high velocity escapers lead to a rapid 
decline in $\dot{M}$ and $\Delta K$ on a timescale of
at most a few $\times 10^5$ yr. The most destructive
encounters considered here (at 10 ${\rm kms}^{-1}$) 
reduce the accretion rate by an order of magnitude in
$10^5$ yr. At lower velocities 
($v_{\rm eject} \approx 2 \ {\rm kms}^{-1}$), the
influence of the interactions tapers off as the
mass and viscous time of the remnant disc approach
the values of the undisturbed system. For the viscosity
and initial conditions (primarily, the choice of initial
disc mass) used here, this occurs at $v_{\rm eject} \ale
2 \ {\rm kms}^{-1}$.

The form of the curves shown in Figure 3 reflects the
relationship between the viscous timescae of the 
undisrupted disc and that of the truncated disc following the
encounter (equation \ref{2.4}). There are little in the
way of observational constraints on the disc viscosity 
at large radius (essentially, just the disc lifetime which 
is expected to be ${\cal{O}}(t_\nu)$ for the disc), and thus
there is bound to be considerable uncertainty inherent
in these timescale estimates. Nonetheless, the results
presented here indicate that encounters close enough to
eject stars at velocities, $v_{\rm eject} \age 5 \ {\rm kms}^{-1}$,
should be disruptive enough of the disc to rapidly cut-off
accretion, and, if the star possesses a magnetosphere
extending to a few stellar radii, additionally lead to
a sharp drop in the near-IR excess. These high velocities
are required if any significant fraction of the 
most dispersed RASS sources are relatively young ($<10$ Myr)
and originated as escapers from known molecular clouds 
(Sterzik et al 1995). Less destructive encounters, with
periastron separations of 20-30 a.u. and ejection
velocities of $\sim 3 \ {\rm kms}^{-1}$, are predicted
to reduce the CTTS phase by modest factors of around 2.

\section{Summary}

In this paper, we have investigated the evolution of discs
around Classical T Tauri stars which are subjected to disruptive
encounters within the environment of small clusters. We find that  
interactions that are able to eject stars from the
cluster at velocities of a few ${\rm kms}^{-1}$ or greater
lead to truncation of the accretion disc, and greatly
accelerate its evolution. The rate of accretion through 
the disc drops rapidly for several viscous times of the
remnant disc, and falls to very low levels within, typically,
a few $\times 10^5$ yr. Since ejections at these velocities
from small clusters happen at an early epoch, the curtailment
of further infall onto the outer regions of discs in escaping
stars would also tend to reduce the accretion rate at later
times.

The further evolution of the discs in ejected systems then
depends on the strength of the stellar magnetic field. For
systems with weak magnetic fields, we predict a period in
which a passive reprocessing disc generates prominent
infra-red excesses at all wavelengths. Alternatively,
for stars with a strong, ordered, magnetic field, the
stellar magnetosphere should rapidly disrupt the inner
regions of the disc out to close to the corotation radius.
As magnetic fields of the required strength are inferred
from other observations, we regard this magnetic scenario
as the more probable. Observationally such systems would show 
negligible infra-red excesses at $K$, but would still possess 
infra-red emission at wavelengths $\lambda \age  5 \ {\mu m}$. 
Such transition systems would have stellar rotation periods 
comparable to the Classical T Tauri stars, and detectable 
emission at mm wavelengths that is, however, depleted by 
the reduction of the disc mass as a result of the encounter.  

In the context of models wherein the Li-rich halo RASS sources 
originate as escapers from young clusters (Sterzik \& Durisen 1995),
the present study implies that the lack of a dispersed 
CTTS population is not a problem, even if some fraction
of the dispersed WTTS turn out to be young ($\sim 1$ Myr). 
The interactions leading to ejection at such high velocities
necessarily cut-off accretion within less than this time.
However there should be systems (preferentially the younger
ones), where residual circumstellar material exists, and
this may be detectable at longer infra-red and mm wavelengths.
Measurements of the proper motions and determinations of
the ages of these RASS sources should provide a decisive
test of such scenarios.

This model, in common with previous work (Clarke et al. 1995), predicts
stars with properties intermediate between Classical and Weak-Lined
T Tauri systems. In particular, if the disc is cleared at the end
of the CTTS phase `inside-out', for example by winds or a stellar
magnetosphere, it is hard to avoid a period in which the system
is `weak' on small scales but `strong' on large scales. Such systems
should have detectable mid infra-red and mm fluxes without 
corresponding $K$ or $H \alpha$ emission. If such systems are not
observed, that would be strong evidence in favour either of
an unexpectedly short disc clearing timescale, or of disc
dissipation via processes unconnected to the central star.

\section*{Ackowledgements}

We would like to thank Michael Sterzik for bringing this problem
to our attention, and for helpful discussions of the
observations. We also thank
the referee, Lee Hartmann, for a candid review of an 
earlier version of this paper. PJA thanks NAOJ, Mitaka, 
for hospitality, and acknowledges financial support 
from the British Council.

\section*{APPENDIX: CALCULATING DISC SPECTRAL ENERGY DISTRIBUTIONS}
			 
The spectral energy distributions for the magnetic disc models
are calculated assuming that annuli
in the disc radiate as blackbodies with effective temperature
$T_{\rm e}$, where
\begin{equation}
	T_{\rm e}^4 = T_\nu^4 + T_{\rm p}^4 + T_{\rm B}^4,
\label{a1}
\end{equation}
and the terms on the right-hand side represent respectively the contributions
from active accretion, reprocessing of the stellar luminosity,
and work done on the disc by the magnetic field. For a disc in which
the energy from viscous dissipation is radiated locally
\begin{equation}
   2 \sigma T_\nu^4 = {9 \over 4} \nu \Sigma \Omega^2,
\label{a2}
\end{equation}
where $\sigma$ is the Stefan-Boltzmann constant. Evaluation of the
right-hand side of this expression does not require knowledge of
the specific form of $\nu$, but for a magnetic disc model it does
depend on the assumed magnetic field configuration. 
In Section 2 we compute
$\nu \Sigma$ from the steady-state solution of the disc evolution
equation (\ref{3.1}) with boundary 
conditions $\nu \Sigma = 0$ at $R = R_{\rm m}$
and $\nu \Sigma = \dot{M} / (3\pi)$ at large radius (the standard
non-magnetic disc expression), yielding
\begin{eqnarray}
   \nu \Sigma = {\dot{M} \over {3 \pi}} \left( 1 - \sqrt{R_{\rm m} \over
   R} \right) - \beta R^{-7/2} \left( \left( R \over R_{\rm m} 
   \right)^3 - 1 \right)  
   \nonumber \\
   + 2 \beta R_{\rm c}^{-3/2} R^{-2} \left(
   \left( R \over R_{\rm m} \right)^{3/2} -1 \right),
\label{a3}
\end{eqnarray}   		 
with $\beta = (B_* R_*^3)^2 / (9 \pi \sqrt{GM_*})$. The main
assumption required to derive this expression is in the form
of the radial dependence of the magnetic torque, and is described
in detail in a previous paper (Armitage \& Clarke 1996). Note that this
solution is not in general physically reasonable ($\nu \Sigma > 0$)
for all choices of $R_{\rm m}$, reflecting the fact that for
large $B_*$ and/or small $\dot{M}$ steady solutions with 
$R_{\rm m} \ll R_{\rm c}$ do not exist. In Section 3 we use
$\nu\Sigma$ calculated directly from the time-dependent solution
of equation (\ref{3.1}).

For the component due to reprocessing of stellar radiation, we utilise
the results of Kenyon \& Hartmann (1987) and Adams \& Shu (1986).
These authors obtain
\begin{equation}
   T_{\rm p}^4 = { T_*^4 \over \pi} \left( \sin^{-1} \left( R_* \over
   R \right) - \left( R_* \over R \right) \sqrt{( 1 - \left( R_* \over
   R \right)^2} \right),
\label{a4}
\end{equation}   
for the case assumed here of a flat reprocessing disc.

The magnetic torques arising from field lines linking the star and the
disc will also do work on the disc material, and some fraction of this
may be thermalized and heat the disc material. Assuming {\em all} the
available energy goes into heating the disc,
\begin{equation}
  2 \sigma T_{\rm B}^4 = (B_* R_*^3)^2 \vert \Omega - \Omega_* \vert
  R^{-5}.
\label{a5}
\end{equation}
The actual fraction of the available energy that goes into reheating
the disc material is uncertain, and so for the purposes of this work we 
adopt the conservative option of choosing the maximum value. Even then,
this term is typically smaller than $T_{\rm p}$, and
so the uncertainty in its value does not seriously change the estimate
of the disc temperature at low accretion rates.

\end{document}